\documentclass[12pt,a4paper]{article}

\usepackage{authblk}
\usepackage{graphicx}
\usepackage{multicol}
\usepackage{epstopdf}
\usepackage{amssymb}
\usepackage{url}
\RequirePackage{mathptmx}
\RequirePackage{flushend}
\RequirePackage[numbers,sort&compress]{natbib}
\RequirePackage[colorlinks,citecolor=blue,urlcolor=blue,linkcolor=blue]{hyperref}

\begin{document}

\title{Investigation on steering of ultrarelativistic $e^{\pm}$ beam through an axially oriented bent crystal}

\author[1]{L.~Bandiera}
\author[2,3]{I.V.~Kyryllin}
\author[4,5]{C.~Brizzolari}
\author[1,6]{R.~Camattari}
\author[7]{N.~Charitonidis}
\author[8]{D.~De~Salvador}
\author[1,6]{V.~Guidi}
\author[4,5]{V.~Mascagna}
\author[1]{A.~Mazzolari}
\author[4,5]{M.~Prest}
\author[1,9]{M.~Romagnoni}
\author[2,3]{N.F.~Shul'ga}
\author[1,6]{M.~Soldani}
\author[1]{A.~Sytov}
\author[5]{E.~Vallazza}
\affil[1]{INFN Sezione di Ferrara, Via Saragat 1, 44122 Ferrara, Italy}
\affil[2]{Akhiezer Institute for Theoretical Physics, National Science Center ``Kharkov Institute of Physics and Technology'', Akademicheskaya Str.,~1, 61108 Kharkiv, Ukraine}
\affil[3]{V.N.~Karazin Kharkiv National University, Svobody Sq. 4, 61022, Kharkiv, Ukraine}
\affil[4]{Universit{\`a} dell'Insubria, via Valleggio 11, 22100 Como, Italy}
\affil[5]{INFN Sezione di Milano Bicocca, Piazza della Scienza 3, 20126 Milano, Italy}
\affil[6]{Dipartimento di Fisica, Universit{\`a} di Ferrara, Via Saragat 1, 44122 Ferrara, Italy}
\affil[7]{CERN, European Organization for Nuclear Research, CH-1211 Geneva 23, Switzerland}
\affil[8]{INFN Laboratori Nazionali di Legnaro, Viale dell'Universit{\`a} 2, 35020 Legnaro, Italy $\&$ Dipartimento di Fisica, Universit{\`a} di Padova, Via Marzolo 8, 35131 Padova, Italy}
\affil[9]{Dipartimento di Fisica, Universit{\`a}  di Milano, Via Giovanni Celoria, 16, 20133 Milano, Italy}

\maketitle

\begin{abstract}
An investigation on stochastic deflection of high-energy charged particles in a bent crystal was carried out. In particular, we investigated the deflection efficiency under axial confinement of both positively and negatively charged particles as a function of the crystal orientation, the choice of the bending plane, and of the charge sign. Analytic estimations and numerical simulations were compared with dedicated experiments at the H4 secondary beam line of SPS North Area, with 120 GeV/$c$ electrons and positrons. In the work presented in this article, the optimal orientations of the plane of bending of the crystal, which allow deflecting the largest number of charged particles using a bent crystal in axial orientation, were found.
\end{abstract}

\section{Introduction}

Due to the fact that bent crystals allow deflecting the direction of motion of high-energy charged particles, these small but powerful objects are of great interest for the HEP and accelerator physics communities, after a number of successful experiments on the deflection of charged particles by bent crystals (see e.g. \cite{biryukov2013crystal,IvanovVolume,afonin2012observation,ScandaleAxial}) and especially after the first demonstration of the possibility to steer the LHC 6.5 TeV/c proton beam through channeling in a compact bent crystal \cite{Scandale2016129}.
Indeed, the strong intra-crystalline field provides the capability of steering an ultra-high-energy beam by bent crystals in a ultra-compact size (mm), impossible to be achieved with state-of-the-art electrostatic or magnetic fields.
This makes bent crystals an excellent candidate for collimating beams in accelerators, deflecting beams, and even splitting them into multiple pieces \cite{splitting1,splitting2}.

If a high-energy charged particle penetrates through a crystal having a small angle $\psi$ between its momentum and one of the main crystallographic axes/planes, significant correlations between successive collisions of the particle with neighboring atoms may occur.
This happens when the angle $\psi$ is of the order of the critical angle of axial channeling \cite{Lin,Akhiezer:Shulga} $\psi_c = \sqrt{4 Z e^2 / p v d}$, where $Z|e|$ is the charge of atomic nucleus in the crystal, $d$ is the distance between atoms in atomic string, $p$ and $v$ are the momentum and velocity of the particle.
In this case, the particle motion is defined by the continuous potential of the atomic strings/planes.

Typically, planar channeling is exploited for the steering of high-energy beams, due to the easier experimental requirements for its implementation \cite{ScandaleQMsps}.
In the case of planar channeling in a bent crystal, the particles must be in under-barrier state, but due to incoherent scattering, some of them dechannel and transit to the regime of over-barrier motion.
Such particles do not deflect through the bending angle of the crystal.
To obtain the axial orientation it is necessary to align the crystal along two directions, but the axial field allows the entire beam to be deflected, which makes it attractive for possible application in beam manipulation.
Only after the advent of new technology for the realization of short bent crystal \cite{Afonin1998,Lanzoni2010} it became possible to achieve the deflection of the whole beam with axially oriented crystal.

In the axial potential, the particle motion in the plane ($x$, $y$) orthogonal to the $z$-axis can be a finite (axial channeling) or an infinite (above-barrier) motion. If the crystal is bent, both axial channeling and above-barrier motion may cause a deflection of the direction of motion of the particles. This second mechanism is usually called stochastic deflection and it is caused by multiple scattering with atomic strings (the so-called \textit{doughnut} scattering) \cite{Shulga1995}. Stochastic deflection was predicted in \cite{Grinenko1991} and allows deflecting the whole beam if the crystal is bent with an angle $\alpha$ \cite{SHULGA2011100}, provided that
\begin{equation}
 \alpha < \alpha_{st}=\frac{2R\psi^2_c}{l_0},
 \label{eq:max}
\end{equation}
where $\alpha_{st}$ is the maximum angle achievable through stochastic deflection, $R$ the radius of crystal curvature, $l_0=\frac{4}{\pi^2ndR_a\psi_c}$, $n$ being the concentration of atoms in the crystal, $d$ the distance between neighboring atoms in the atomic string, and $R_a$ the atomic screening radius. Condition (\ref{eq:max}) was found without taking into account incoherent scattering, i.e. $\alpha_{st}$ is a function of the particle energy and $R$, but the thickness of the crystal is not included in Eq. (\ref{eq:max}) as a parameter.

In the pioneering experiments on axial channeling \cite{Bak1984,Baurichter1986} an efficient beam deflection at the nominal bending angle was not observed because Eq. (\ref{eq:max}) was not fulfilled. In these experiments, stochastic deflection took place only at small crystal thicknesses, for which condition (\ref{eq:max}) was fulfilled. At larger crystal thicknesses, the particles passed into the mode of planar channeling in non-vertical atomic planes and, as a result, were deflected at angles that were much smaller than the bending angle of the crystal. In the following experiments stochastic deflection was experimentally observed for both positively \cite{ScandaleAxial,splitting2,SCANDALE2016826,splitting1,garattini-epjc} and negatively \cite{Scandale2009301} charged hadrons, owing to the advent of a new generation of bent crystals \cite{Afonin1998,Lanzoni2010,GERMOGLI201581}, short enough to efficiently deflect high-energy particles up to the nominal bending angle of the crystal.

In this paper, we present an investigation through experiments, analytical estimations, and Monte Carlo simulations on the axial deflection of $e^{\pm}$ beams in bent crystals and its dependence on crystal curvature and choice of crystal axis and bending planes. This is indeed the first experimental evidence of axial stochastic deflection of ultrarelativistic leptons in bent crystals, since only planar channeling and volume reflection were investigated for high-energy electrons and positrons \cite{Ban3,PhysRevLett.112.135503,SYTOV2017,Bagli2017,Lie,Wistisen2017,PhysRevLett.114.074801,wistisen2016channeling,ban}. With this work, we also addressed possible schemes for manipulation of $e^{\pm}$ beams, such as steering or extraction, in present and future electron/positron (and in general lepton) accelerators and colliders.

\section{Negatively charged particles}

In \cite{Kirillin}, on the basis of an analytic calculation and a numerical simulation of the motion of high-energy negatively charged particle in a bent crystal, the existence of the optimal radius of crystal curvature was shown. The optimal radius of curvature is the radius for which most of the beam is deflected at a maximal angle through the stochastic deflection mechanism. Let us now compare the efficiency of the stochastic deflection for the two main crystallographic axes of a silicon crystal, i.e. $\langle 110 \rangle$ (the strongest one) and $\langle 111 \rangle$. The choice of silicon is due to its high degree of crystallographic quality, close to perfection, and its relative low price. Indeed, such material has been chosen for the application of crystals as primary collimators of the LHC hadron beam \cite{Scandale2016129}.

Since condition (\ref{eq:max}) was found without taking into account incoherent scattering, which is quite important for negatively charged particles that move close to the high atomic density region of the crystal axes being attracted by the positive atomic nuclei, in \cite{Kirillin} it was shown that the dependence of the crystal thickness $L_{st}$, up to which negatively charged particles take part in the stochastic deflection, on the radius of curvature $R$ has the following form
\begin{equation}
	\label{eq:Lst}
	L_{st} = \frac {\psi_{m}^{2}} {l / R^{2} + \xi},
\end{equation}
where $\psi_{m}$ is the maximum value of the angle $\psi$ for which particles take part in the stochastic deflection, $l$ is the mean length of the path that the particle crosses during scattering on one atomic string, and $\xi$ is a constant of proportionality between the mean square angle of incoherent multiple scattering on atomic thermal vibrations, electronic subsystem atoms, etc., and the thickness of the crystal. From Eq. (\ref{eq:Lst}), we see that the stochastic deflection mechanism gives the possibility to deflect a beam of negative particles up to the maximum angle
\begin{equation}
	\alpha_{st} = \frac {L_{st}}{R} = \frac {\psi_{m}^{2}} {l / R + \xi R}.
	\label{eq:alp}
\end{equation}
The dependence of this angle on the particle energy is shown in \cite{kirillin_dependence_2017-1}, where it is shown that with increasing energy $\alpha_{st}$ decreases more slowly than the critical angle of axial channeling.

The angle $\psi_m$ for negatively charged particles is about $1.5\psi_c$ \cite{Kirillin}. Thus, in order to compare $\alpha_{st}$ for the $\langle 110 \rangle$ and $\langle 111 \rangle$ axes of Si crystal, one needs to compare the value of $\psi_c^2$ for these axes: $\left(\psi_c^{\langle 110 \rangle} / \psi_c^{\langle 111 \rangle}\right)^2 \approx 1.22$, so in case of crystal orientation along the $\langle 110 \rangle$ axis, stochastic deflection of negative particles should be more effective with respect to the case of orientation along the $\langle 111 \rangle$ axis.

To prove the last statement, which follows from analytical calculations, we carried out a numerical simulation of the motion of negatively charged particles in a bent crystal in conditions of stochastic deflection.
For definiteness and for comparison with experiments presented in Section 4, the momentum of the particles was chosen to be 120 GeV/$c$; the charged particles were electrons.
The number of particles that impinged on the crystal was 2$\times 10^3$.
For the selected energy, the critical axial channeling angle for electrons and positrons is 41.8 $\mu$rad for axis $\langle 110 \rangle$ and 37.8 $\mu$rad for axis $\langle 111 \rangle$.

\begin{figure}[h]
\includegraphics[width = \columnwidth]{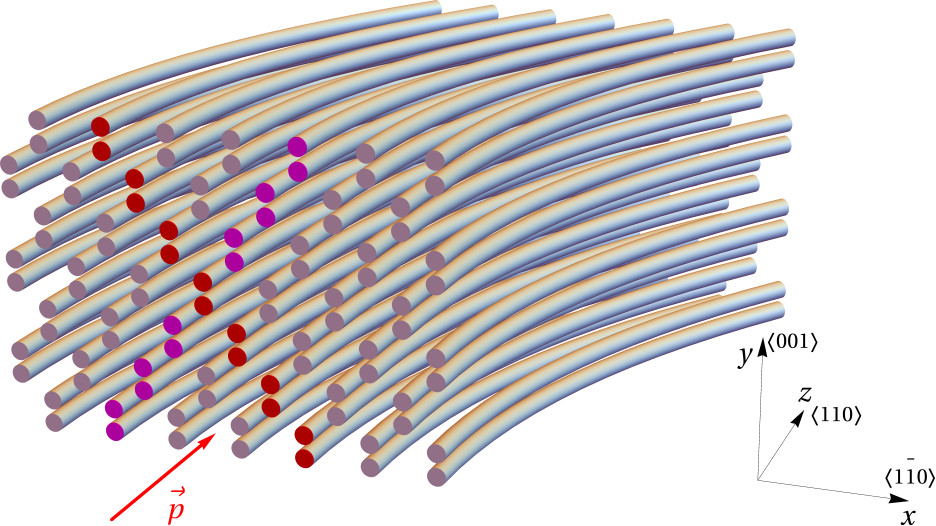}
\includegraphics[width = 0.85\columnwidth]{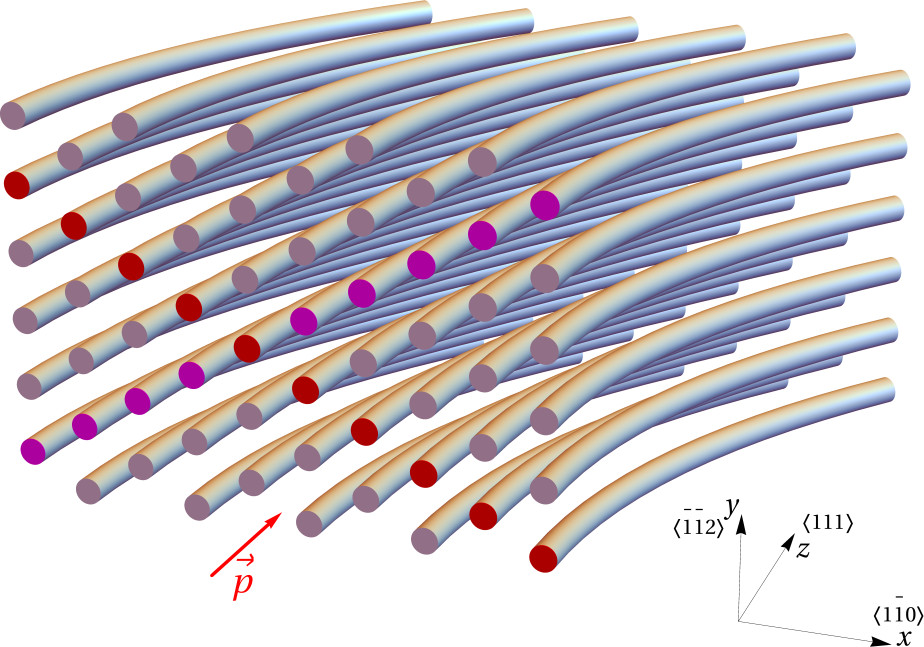}
\caption{Illustration of atomic strings location in Si crystal oriented along the $\langle 110 \rangle$ axis (top figure) and along the $\langle 111 \rangle$ axis (bottom figure). For the $\langle 110 \rangle$ orientation red color shows $(1\bar{1}1)$ plane, magenta color shows $(\bar{1}11)$ plane and for the $\langle 111 \rangle$ orientation red color shows $(0\bar{1}1)$ plane, magenta color shows $(\bar{1}01)$ plane.}
\label{fig:illustration}
\end{figure}

In the simulation, the bending plane of the crystal coincided with the (001) crystal plane in case of orientation along the $\langle 110 \rangle$ axis and with the $(\bar{1}\bar{1}2)$ crystal plane in case of orientation along the $\langle 111 \rangle$ axis, as shown in Fig. \ref{fig:illustration}. The bent vertical plane is the one orthogonal to the bending plane, being $(1\bar{1}0)$ atomic plane in both cases. Beam divergence in the simulation was set equal to zero. The Monte Carlo code was the same as in Ref. \cite{SHULGA2011100,Relaxation,optimal}. The code solves the two-dimensional equation of motion in the field of continuum atomic string potential through numerical integration and also takes into account the contribution of incoherent scattering with atomic nuclei and electrons. Other incoherent effects were not taken into account due to the small thickness of the crystal.
In particular, the effects associated with radiation energy losses were not taken into account, since the crystal thickness was much smaller than the radiation length.

\begin{figure}[h]
\includegraphics[width = \columnwidth]{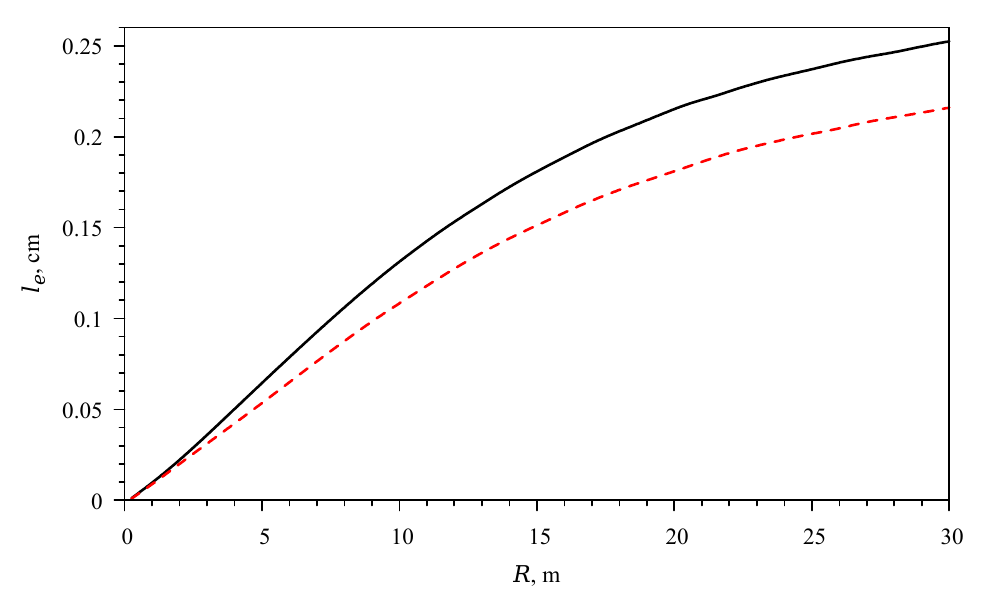}
\caption{Dependence of the relaxation length as a function of the radius of curvature for the crystal orientation along the $\langle 110 \rangle$ axis (black solid line) and along the $\langle 111 \rangle$ axis (red dashed line) with respect to the impinging particles.}
\label{fig:le}
\end{figure}

In Fig.\ref{fig:le}, we show the dependence of the crystal length $l_e$ (the relaxation length), i.e. the length within which the number of electrons that moves in the crystal in stochastic deflection regime decreases by a factor of $e$ \cite{Relaxation}, as a function of the crystal radius of curvature. The black solid line corresponds to the motion in a crystal oriented along the $\langle 110 \rangle$ axis, while the red dashed line corresponds to motion in a crystal oriented along the $\langle 111 \rangle$ axis. One can see that for small bending radii, the relaxation length grows fast with $R$, while for larger radii the speed of growth of $l_{e}$ decreases. For the given $R$, the relaxation length in case of $\langle 110 \rangle$ orientation is higher than in case of $\langle 111 \rangle$ orientation, but the difference between $l_e$ depends on $R$ (as for $R \to 0$ for any orientation $l_e \to 0$).

\begin{figure}[h]
\includegraphics[width = \columnwidth]{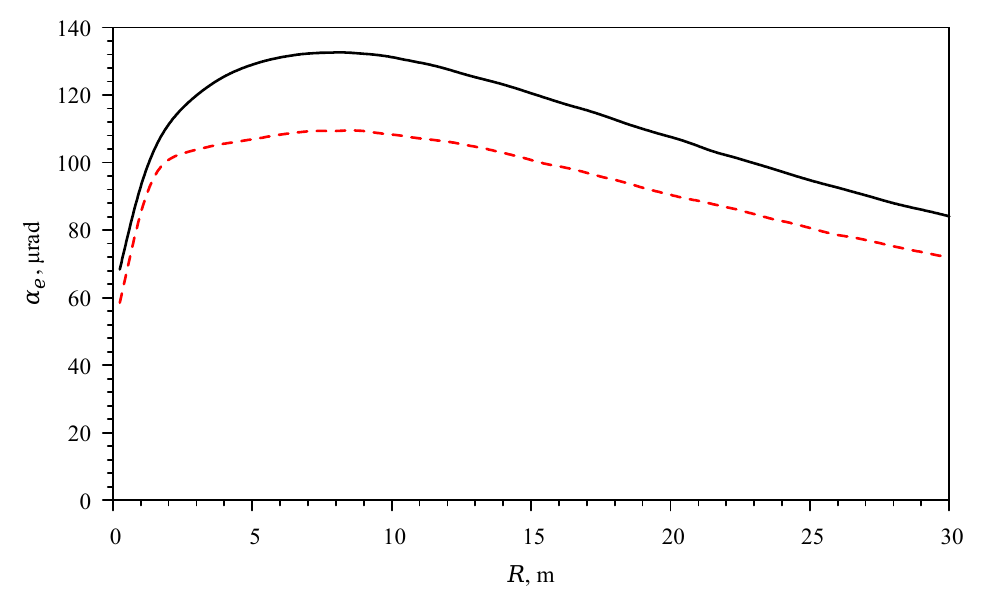}
\caption{Dependence of the deflection angle $\alpha_e$ on the radius of curvature for the crystal orientation along the $\langle 110 \rangle$ axis (black solid line) and along the  $\langle 111 \rangle$ axis (red dashed line) with respect to the impinging particles.}
\label{fig:alpe}
\end{figure}

In order to find the optimal value for the radius of curvature, we analyzed via simulation the dependence of the deflection angle, at which $\frac{1}{e}$-th part of beam particles is deflected, over the radius of curvature. The dependence of this deflection angle, $\alpha_{e} = l_{e} / R$, from the radius of curvature is plotted in Fig. \ref{fig:alpe}. As in Fig. \ref{fig:le}, the black solid line corresponds to the $\langle 110 \rangle$ orientation, while the red dashed line corresponds to the $\langle 111 \rangle$ orientation.
In Fig. \ref{fig:alpe} we see that for both orientations the value of the optimal radius of curvature is about 8 m. One can see that at $R = R_{opt}$ the value of $\alpha_{e}$ for the $\langle 110 \rangle$ orientation is about 1.2 times higher than for the $\langle 111 \rangle$ orientation (as it should be, according to Eq. (\ref{eq:alp})).

\begin{figure}[h]
\includegraphics[width = \columnwidth]{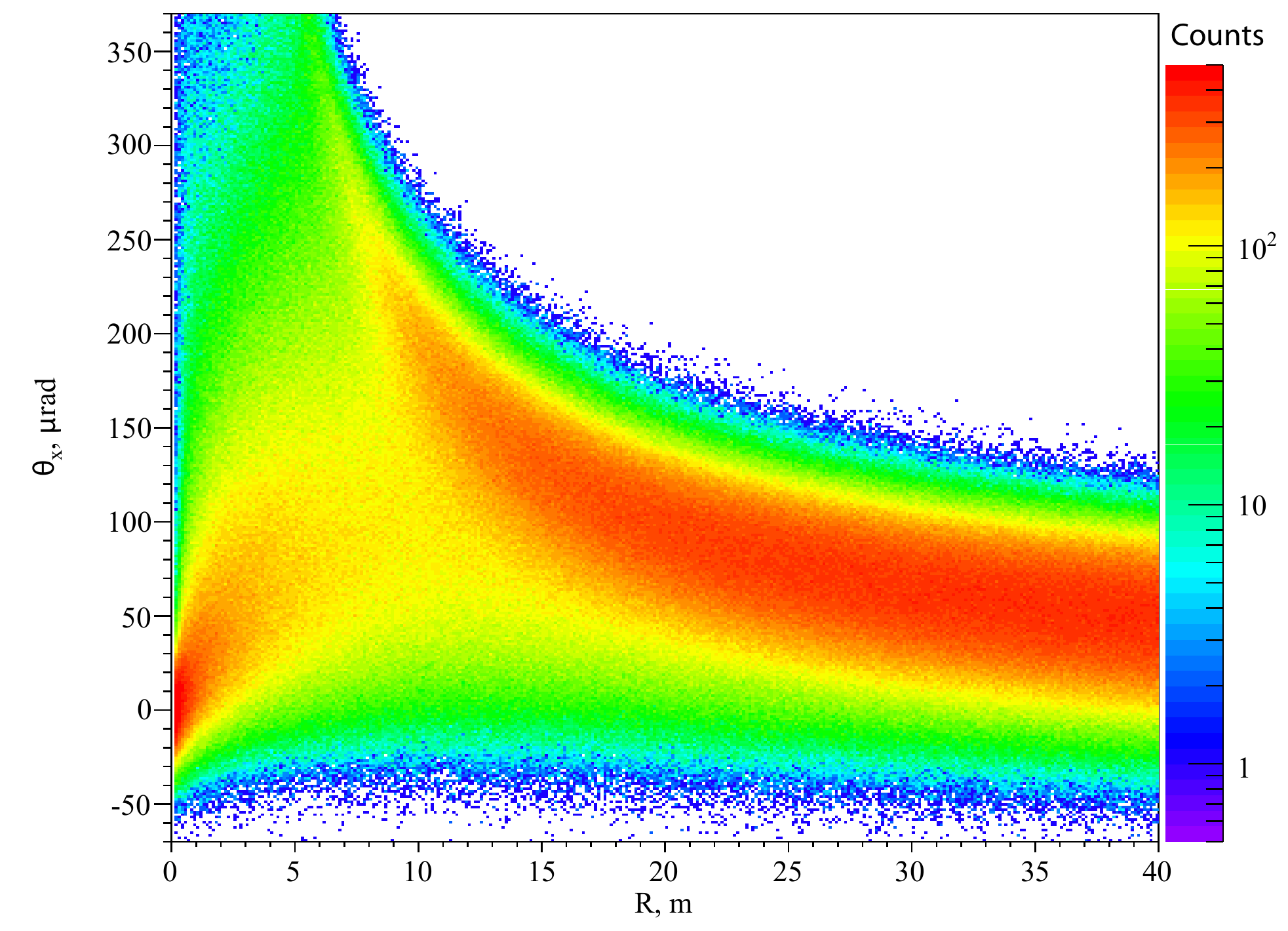}
\caption{The dependence of the number of electrons deflected on the angle $\theta_x$ on the radius of curvature of the crystal for the $\langle 110 \rangle$ crystal orientation with respect to impinging particles.}
\label{fig:prof110}
\end{figure}

\begin{figure}[h]
\includegraphics[width = \columnwidth]{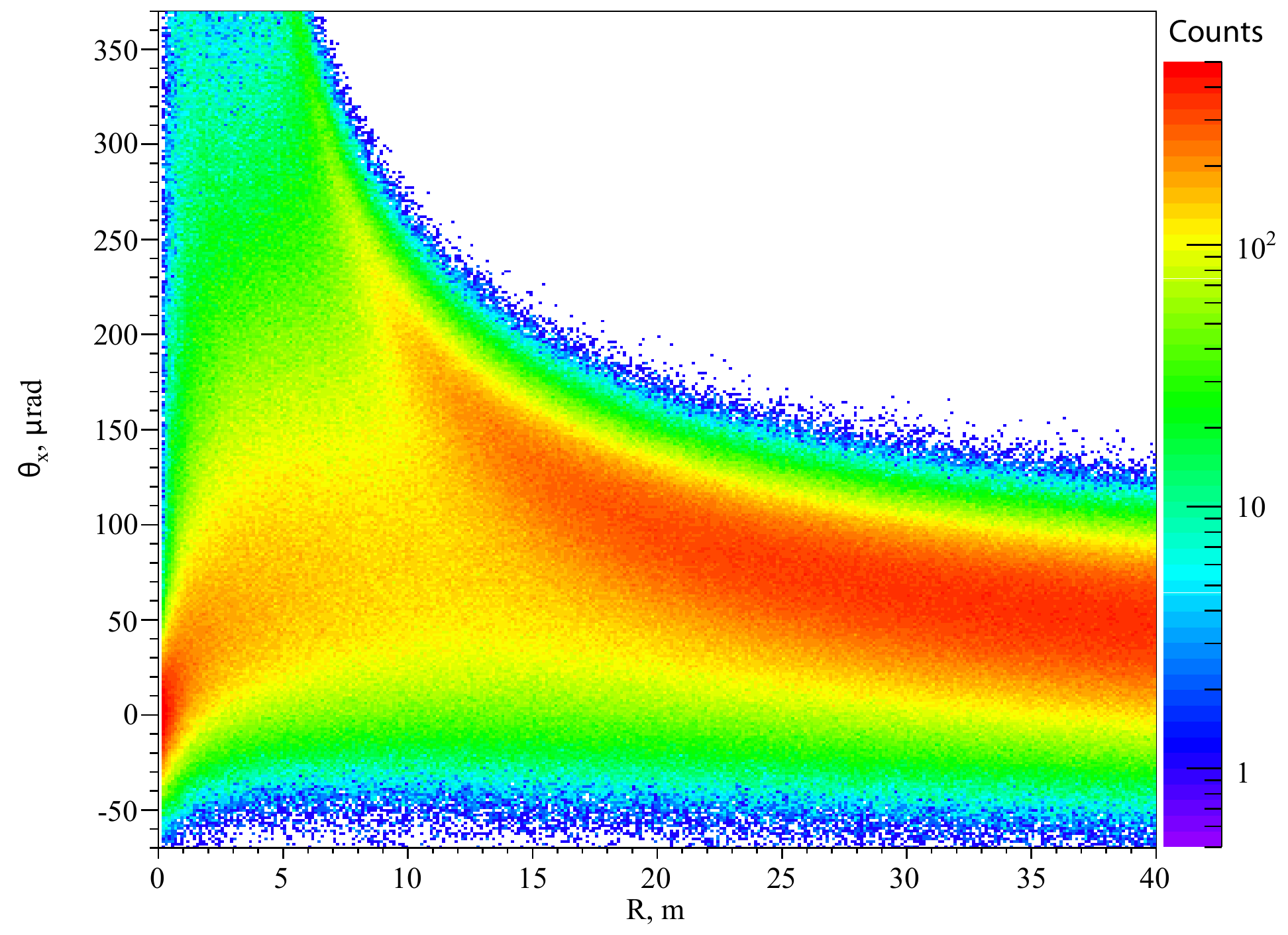}
\caption{The dependence of the number of electrons deflected on the angle $\theta_x$ on the radius of curvature of the crystal for $\langle 111 \rangle$ crystal orientation with respect to impinging particles.}
\label{fig:prof111}
\end{figure}

In Fig. \ref{fig:prof110} and Fig. \ref{fig:prof111} we also show the dependence of the number of particles, deflected at an angle $\theta_x$, as a function of the crystal radius of curvature for $\langle 110 \rangle$ and $\langle 111 \rangle$ orientations, respectively. The thickness of the crystal in the simulation was 2 mm (as in the experiments reported in Section 4). Colors show the number of particles deflected at the given angle $\theta_x$. We see that the dependencies in Fig. \ref{fig:prof110} and Fig. \ref{fig:prof111} are similar, but in the case of the $\langle 111 \rangle$ orientation the angular distribution of the beam after crossing the crystal is a bit wider. Moreover, in the case of the $\langle 110 \rangle$ orientation, a larger number of particles is deflected to the angle of curvature of the crystal.
It can be seen that the larger the deflection angle (or smaller the bending radius), the less particles deflect by this angle.
Thus, for those applications in which it is necessary to deflect the majority of the particles (and not only the $1/e$-th part of the particles) at a given angle, a bent crystal with $R > R_{opt}$ may be more acceptable.

\section{Positively charged particles}

Positively charged particles crossing an oriented crystal spend less time near atomic strings than negatively charged particles, due to the repulsive force between positive particles and atomic nuclei.
That is why the coefficient $\xi$ from Eq. (\ref{eq:Lst}) for positive particles is much smaller than for negative ones\footnote{The mean square angle of incoherent multiple scattering in crystal depends on the probability of close collisions. The dependence of this probability on crystal orientation in a bent crystal was described in \cite{chesnokov_about_2014,kirillin_orientation_2015,kirillin_dependence_2017}.}.
As a consequence, the dependence of the relaxation length for positive particles as a function of the crystal radius of curvature is close to a parabolic function \cite{Relaxation}.
Therefore, the higher the radius of curvature, the higher the deflection angle it is possible to obtain with a help of stochastic deflection.

Let us carry out the comparison of the efficiency of stochastic deflection of positively charged particles in a Si crystal oriented along the $\langle 110 \rangle$ axis and along the $\langle 111 \rangle$ axis. As in the case of negative particles, which was considered in the previous section, for definiteness, the momentum of the particles was chosen to be 120 GeV/$c$. The charged particles were positrons.

\begin{figure}[!h]
\includegraphics[width = \columnwidth]{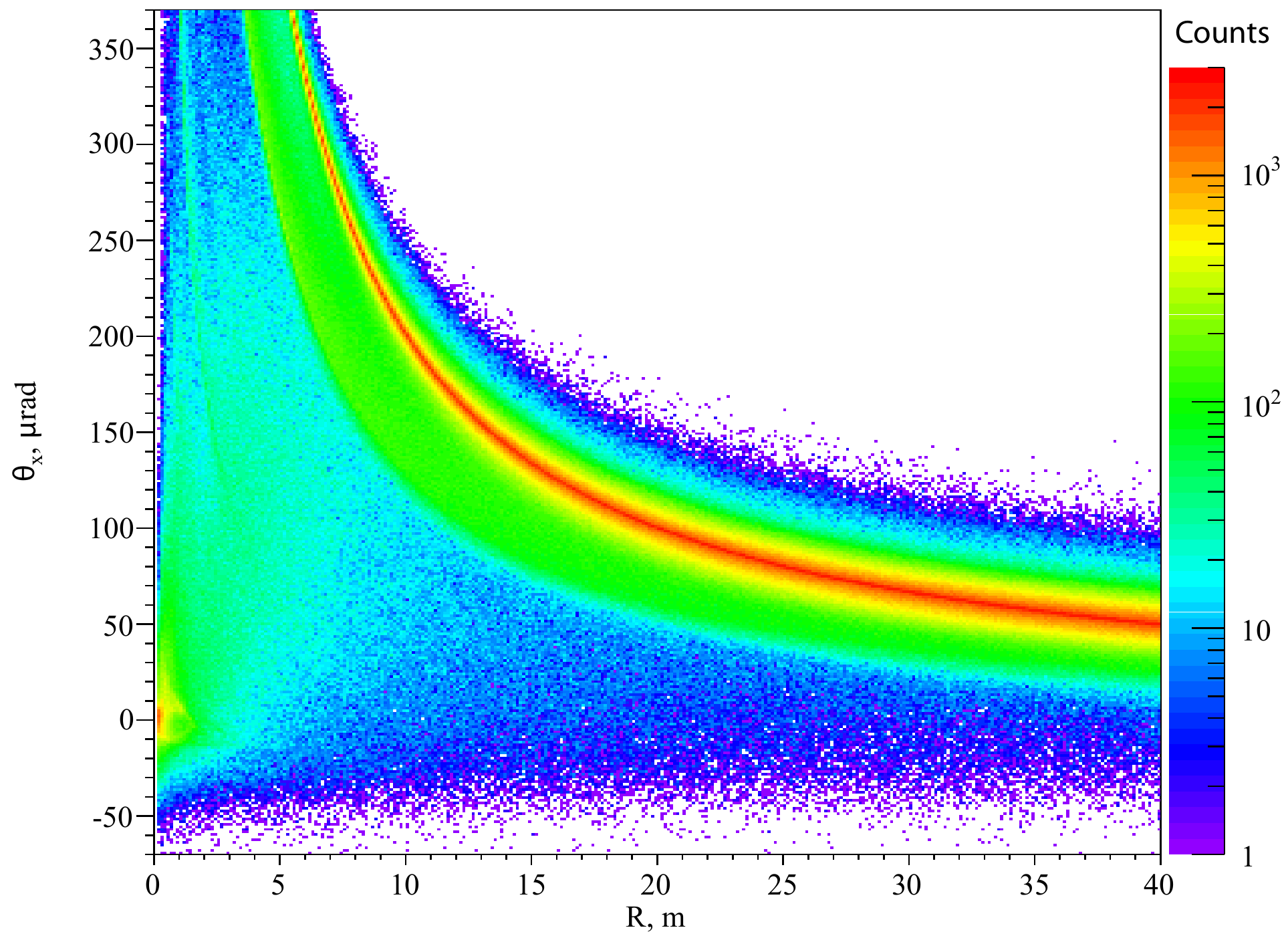}
\caption{The dependence of the number of positrons deflected on the angle $\theta_x$ on the radius of curvature of the crystal for the $\langle 110 \rangle$ crystal orientation with respect to impinging particles.}
\label{fig:prof110-pos}
\end{figure}

\begin{figure}
\includegraphics[width = \columnwidth]{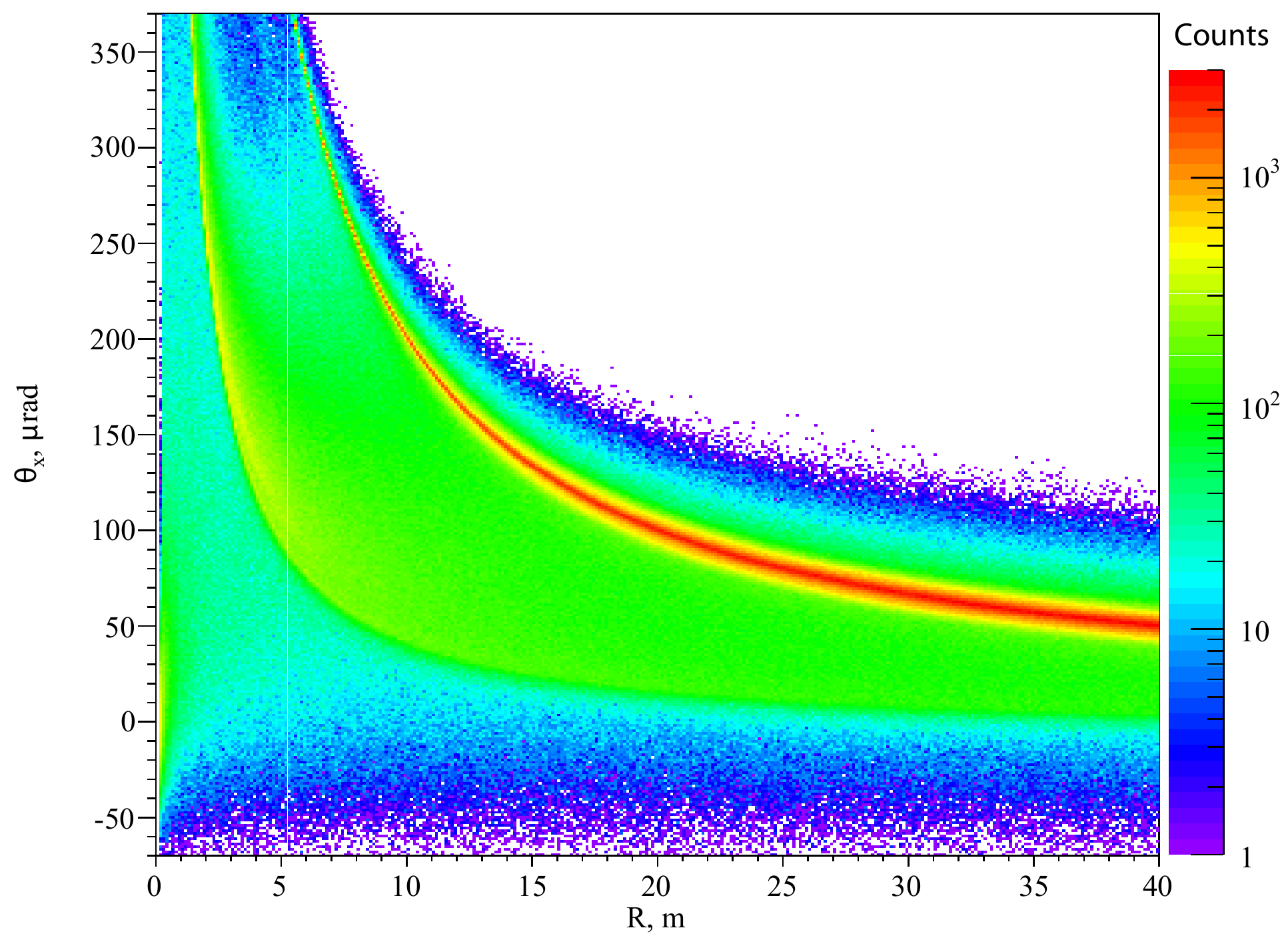}
\caption{The dependence of the number of positrons deflected on the angle $\theta_x$ on the radius of curvature of the crystal for the $\langle 111 \rangle$ crystal orientation with respect to impinging particles.}
\label{fig:prof111-pos}
\end{figure}

In Fig. \ref{fig:prof110-pos} and Fig. \ref{fig:prof111-pos} we present the simulation results with the dependence of the number of particles deflected at an angle $\theta_x$ as a function of the crystal radius of curvature for $\langle 110 \rangle$ and $\langle 111 \rangle$ orientations, respectively. The thickness of the crystal in the simulation was 2 mm. Colors show the number of particles deflected on the given angle $\theta_x$. We see that, unlike in the case of negative particles, the dependencies in Fig. \ref{fig:prof110-pos} and Fig. \ref{fig:prof111-pos} have different features. Indeed, the angular distribution along $x$ axis for the $\langle 111 \rangle$ orientation is much wider that in case of the $\langle 110 \rangle$ orientation. This happens because in the case of positive particles, the main reason of relaxation from stochastic deflection is the capture in planar channeling in skew planes \cite{Relaxation} (see Fig. \ref{fig:illustration}). Indeed, when a charged particle moves in the regime of stochastic deflection, it experiences the action of the centrifugal force that is directed opposite to the curvature vector. The smaller is the angle $\phi$ between the skew plane and the bending plane, the higher is the projection of the centrifugal force to the direction of the skew plane (this projection is proportional to $\cos\phi$). For the $\langle 111 \rangle$ orientation, the angle between the main skew planes, $(0\bar{1}1)$ and $(\bar{1}01)$, and the bending plane $(\bar{1}\bar{1}2)$ is 30$^\circ$, while for the $\langle 110 \rangle$ orientation, the angle between the main skew planes, $(1\bar{1}1)$ and $(\bar{1}11)$, and the bending plane $(001)$ is $\approx 54.7^\circ$ (see illustration of atomic strings location in Fig. \ref{fig:illustration}, where for the $\langle 110 \rangle$ orientation red color shows $(1\bar{1}1)$ plane, magenta color shows $(\bar{1}11)$ plane and for the $\langle 111 \rangle$ orientation red color shows $(0\bar{1}1)$ plane, magenta color shows $(\bar{1}01)$ plane). As shown in figures \ref{fig:prof110-pos} and \ref{fig:prof111-pos}, for the same $R$ more particles stay in the stochastic deflection regime up to the end of the crystal in the $\langle 110 \rangle$ orientation. Moreover, because of smaller $\phi$, for the $\langle 111 \rangle$ orientation, positrons that relax from stochastic deflection to planar channeling in skew planes are deflected at a smaller angle in comparison to the case of the $\langle 110 \rangle$ orientation. This gives us the possibility to state that stochastic deflection in case of positive particles is more effective for the $\langle 110 \rangle$ orientation.
Figures \ref{fig:prof110-pos} and \ref{fig:prof111-pos} do not show the presence of an optimal crystal curvature radius, since in the case of positively charged particles, incoherent scattering is much less intense than for negatively charged particles, however, at $R \to 0$, due to the strong centrifugal force, the particles leave the stochastic deflection mode immediately after entering the crystal.

\section{Experiment at the SPS H4 line with 120 GeV/$c$ $e^{\pm}$}

We tested for the first time the stochastic deflection mechanism using electrons and positrons with an experiment carried out in the H4 beam line of the CERN Super Proton Synchrotron (SPS) - North Area \cite{NA-beam}, where a quite pure electron (or positrons) beam of 120 GeV/$c$ was available.
The beams are slightly contaminated by the presence of hadrons: less than 1\% for $e^-$ and around 10\% for $e^+$.
The hadronic contamination is mainly composed of $p$, $\pi$ and $K$, that provide basically the same critical angle of axial channeling as for $e^{\pm}$ at these ultrarelativistic energies.
The beam intensity (defined by the collimators of the line, located upstream the experimental location) was a few kHz over an effective spill length of 4.8 seconds.
The final focusing quadrupoles of the H4 line were configured to provide a parallel beam, with a divergence of about 90 $\mu$rad in both horizontal and vertical planes.
The change between the electrons and the positrons was very easy, by loading a pre-configured file from the beam control system.

\begin{figure}[h]
\centering
\includegraphics[width=0.9\columnwidth]{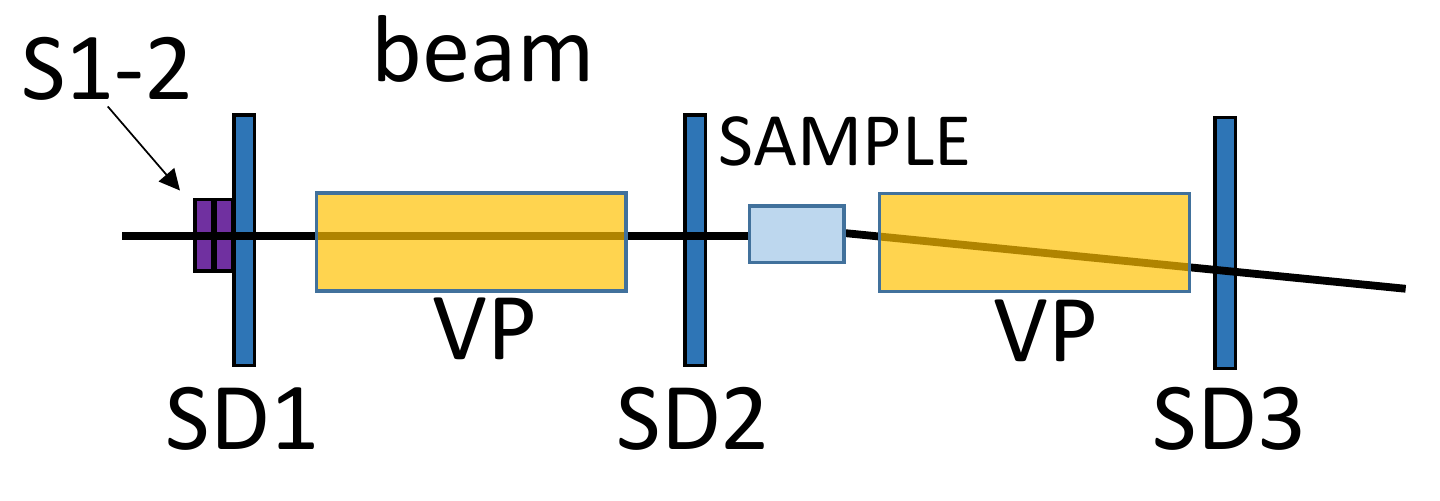}
\caption{\label{fig:setup}Experimental setup. S1-2 are the plastic scintillators used for the trigger to select charged particles; SD1, SD2, and SD3 are the Silicon Detectors forming the tracker. The solid line represents the deflected beam.
For the measurements with the Si crystal bent along the $\langle 110 \rangle$ axis, the SD2, the sample, and the SD3 were placed at 5168 mm, 5821 mm and 11521 mm from the SD1 detector, respectively. For the Si crystal bent along the $\langle 111 \rangle$ axis, SD2 was placed at 5207 mm, the crystal at 5567 mm and the SD3 at 11497 mm from the SD1. Between SD1 and SD2 and between the crystal and SD3 vacuum pipes (VP) were placed to minimize the multiple scattering.
}
\end{figure}

\begin{figure}[h]
  \centering
	\includegraphics[width = 1\columnwidth]{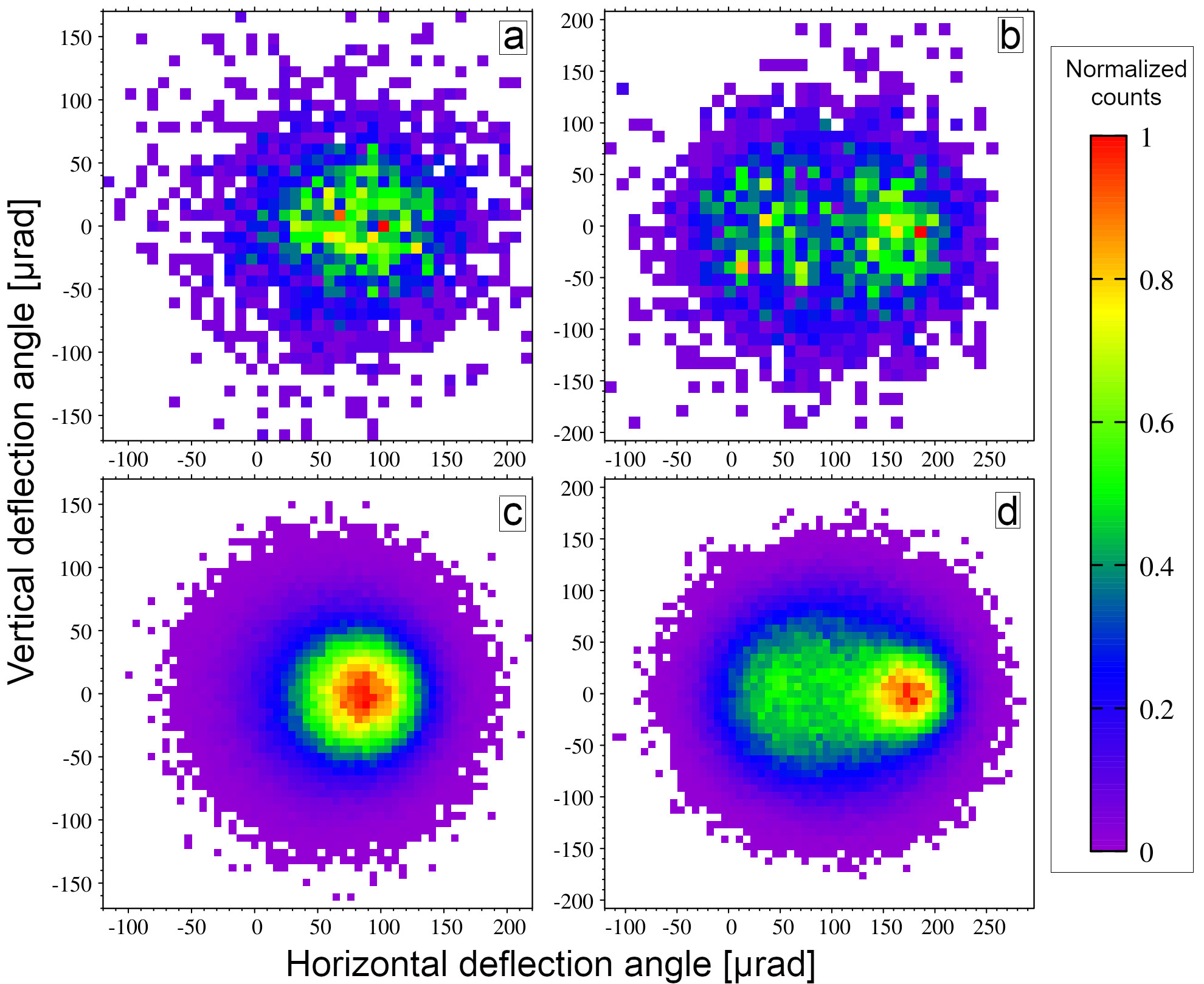}
\caption{Experimental ({\bf a},{\bf b}) and simulated ({\bf c},{\bf d}) angular distribution of 120 GeV/$c$ electrons after passage through 2~mm long Si crystals, aligned with the $\langle 110 \rangle$ ({\bf a},{\bf c}) and $\langle 111 \rangle$ ({\bf b},{\bf d}) axis, bent with curvature radii equal to 22.22 m ({\bf a},{\bf c}) and 11.11 m ({\bf b},{\bf d}).}
	\label{fig:expel}
\end{figure}

For this experiment, two Si crystals ({\it L} = 2 mm thick along the beam direction) were fabricated and oriented along two different axes: $\langle 110 \rangle$ and $\langle 111\rangle$. The crystals were fabricated according to the procedure described in Ref. \cite{Baricordi2007,baricordi2008} and bent through anticlastic deformation \cite{Lanzoni2010}. The $\langle 110 \rangle$ strip was shaped as a parallelepiped of size 2$\times$55.0$\times$2 mm$^3$ and was bent in the $(\bar{1}12)$ bending plane.
The bending angle of the $(1\bar{1}1)$ plane (which was orthogonal to the bending plane) was about $90 \pm 5$ $\mu$rad, as measured through planar channeling of positrons.
The $\langle 111 \rangle$ strip was shaped as a parallelepiped of size 2$\times$55.0$\times$2 mm$^3$ and was bent in the $(\bar{1}\bar{1}2)$ bending plane.
The bending angle of the $(1\bar{1}0)$ plane (which was orthogonal to the bending plane) was about $180 \pm 5$ $\mu$rad.
As a consequence of mechanical imperfections in the bending device, crystals were characterized by torsion ($40 \pm 1$ $\mu$rad/mm for the crystal oriented along the $\langle 110 \rangle$ axis and $11 \pm 1$ $\mu$rad/mm for the crystal oriented along the $\langle 111 \rangle$ axis).
Negative effect of this unwanted deformation on steering efficiency was removed during the offline data analysis as in \cite{ROSSI2015369}. 
For both crystals the bending angles were defined also with high-resolution X-ray diffraction and we found them in agreement with the measurement on the beam.

A sketch of the setup is presented in Fig.~\ref{fig:setup}.
A similar setup is also described in \cite{ban}.
The crystal was mounted on a high-precision goniometer with the possibility to be aligned in both the horizontal and vertical directions, with an accuracy of 1 $\mu$rad in both directions.
The trajectories of the particles interacting with the crystals have been reconstructed with a telescope system based on high-precision micro strip detectors \cite{Lietti2013} (SD1, SD2, and SD3 in Fig.\ref{fig:setup}).
The usage of 3 detectors minimize the multiple scattering contribution to the angular resolution of the telescope \cite{telescope}.
The incoming angle was measured with SD1-SD2 detectors, while the outgoing angle with the SD3 measurement with respect to the estimated position of the particle at the entry face of the crystal.
The angular resolution of the incoming arm (SD1-SD2) was computed through an approach already developed in \cite{telescope} taking into account the intrinsic resolution of the detectors (about 6.4 $\mu$m and 10.1 $\mu$m, for horizontal and vertical incoming angles), and the multiple scattering in the material/air on the beamline.
The total angular resolution for the incidence angle was dominated by the multiple scattering in the SD1 (300 $\mu$m of Si) and was about 7 $\mu$rad, for both directions.
The total telescope angular resolution in the deflection angle was measured directly as the standard deviation of the deflected beam taken with the crystal out of the beam, and was estimated to be about 14 $\mu$rad.
This value is in  agreement with Geant 4 simulation \cite{Bagli2017independence}.
\begin{figure}
  \centering
	\includegraphics[width = 1\columnwidth]{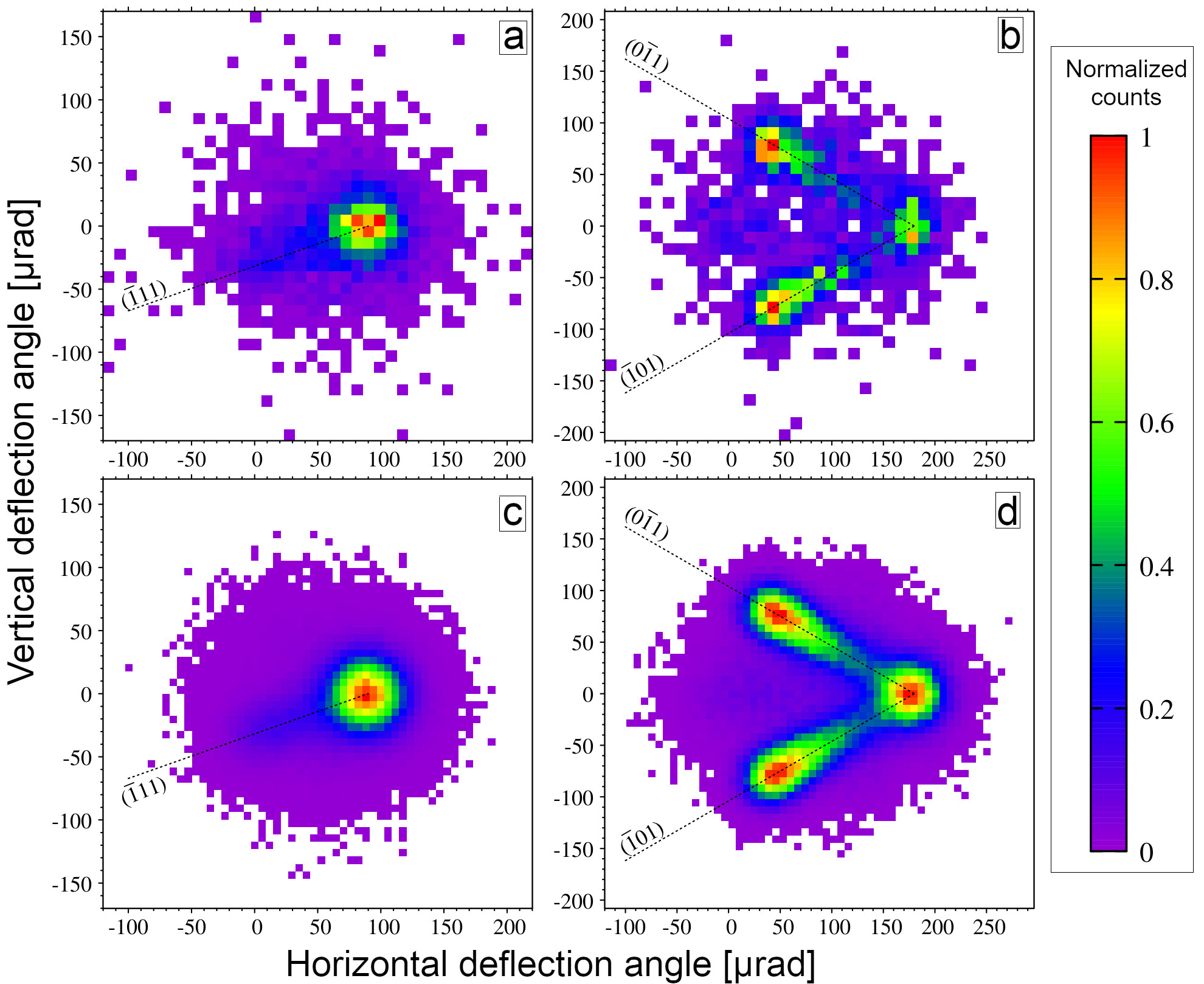}
	\caption{Experimental ({\bf a},{\bf b}) and simulated ({\bf c},{\bf d}) angular distribution of 120 GeV/$c$ positrons after passage through 2~mm long Si crystals, aligned with the $\langle 110 \rangle$ ({\bf a},{\bf c}) and  $\langle 111 \rangle$ ({\bf b},{\bf d}) axis, bent with curvature radii equal to 22.22 m ({\bf a},{\bf c}) and 11.11 m ({\bf b},{\bf d}). The black dotted lines identify the skew planes directions.}
	\label{fig:exppos}
\end{figure}

By exploiting the horizontal rotational movement of the goniometer, we first attained planar channeling in the $(1\bar{1}1)$ planes, then by scanning the vertical rotational movement the crystal was aligned with the $\left< 110 \right>$ axis. An angular cut satisfying the condition $\sqrt{\theta_{X,in}^2+\theta_{Y,in}^2}\leq$ 10 $\mu$rad, $\theta_{X,in}$ and $\theta_{Y,in}$ being the horizontal and vertical incidence angles respectively, was set in order to select only the particles aligned with the $\left< 110 \right>$ crystal axes. A similar procedure was carried out for the crystal alignment along the $\left< 111 \right>$ axis.

The experimental results are shown in Fig. \ref{fig:expel} for the electron beam, while Fig. \ref{fig:exppos} regards the positron beam.
Colors show the number of particles. (a,b) correspond to the experimental results. The figures also show the simulation results for the $\langle 110 \rangle$ (c) and $\langle 111 \rangle$ (d) axes. To reproduce the experimental results, the radius of curvature of the crystal was chosen to be 22.22 m for $\langle 110 \rangle$ and 11.11 m for $\langle 111 \rangle$ axes.
The experimental incoming and deflection angle resolution was also taken into account.

Even with a not very high statistics of the experimental data, the comparison with the simulation highlights a fairly good agreement, which confirms our calculation and assumption on stochastic deflection vs. different crystal axes and radii of curvature, and thus could be exploited to investigate the best choice of these parameter. From experimental data it is indeed clear that the deflection efficiency is more pronounced for the $\langle 110 \rangle$ than for the $\langle 111 \rangle$ axes, since in the latter case the radius of curvature was not optimal to allow the stochastic deflection of the whole beam.
We may notice that for deflection of positive particles the expected relaxation into skew planar channels \cite{Relaxation} is asymmetric in the case of the $\langle 110 \rangle$ axis, for which relaxed particles go mainly in one skew $(\bar{1}11)$ plane (this is due to the choice of bending plane which in the experiment was the $(\bar{1}12)$ plane), and instead symmetric for the $\langle 111 \rangle$ axis (due to the presence of two strong skew planes, $(0\bar{1}1)$ and $(\bar{1}01)$).
Because of symmetric location of two strong skew planes w.r.t. bending plane in case of the $\langle 111 \rangle$ axis in both experimental and simulation results we see three highly populated angular areas in Fig. \ref{fig:exppos}(b,d).
The first is near bent atomic string direction at the end of the crystal ($\theta_x = 180$ $\mu$rad, $\theta_y = 0$).
This angular region contains particles that move in the stochastic deflection mode until the end of the crystal.
The second and the third areas contain particles which relaxed from stochastic deflection mode to planar channeling in skew planes $(0\bar{1}1)$ and $(\bar{1}01)$.
These particles are deflected in both $x$ and $y$ directions and since they relax from stochastic deflection mode not simultaneously, but gradually, these particles are located in Fig. \ref{fig:exppos}(b,d) along two segments: the first from point $\theta_x = 180$ $\mu$rad, $\theta_y = 0$ to $\theta_x = 45$ $\mu$rad, $\theta_y = 45\sqrt{3}$ $\mu$rad (planar channeling in the field of planes $(0\bar{1}1)$) and the second from point $\theta_x = 180$ $\mu$rad, $\theta_y = 0$ to $\theta_x = 45$ $\mu$rad, $\theta_y = -45\sqrt{3}$ $\mu$rad (planar channeling in the field of planes $(\bar{1}01)$).
For electrons, both from the experimental results and from the simulation results, we obtained the number of particles deflected in the direction of the crystal bending (at an angle larger than zero) by more than 91\% for the $\langle 110 \rangle$ axis and more than 89\% for the $\langle 111 \rangle$ axis, while for positrons this number exceeded more than 93\% for both crystal axes.

In sections 2 and 3 it was assumed that the bent crystal with orientation $\langle 110 \rangle$ was bent in the (001) bending plane, however, in section 4 the crystal with orientation $\langle 110 \rangle$ was bent in $(\bar{1}12)$ bending plane.
The dependence of the deflection efficiency on the bending planes was never discussed before, but it is of clear interest to determine the better choice for future applications in beam steering/\hspace{0cm}collimation/\hspace{0cm}extraction from particle accelerators.

\section{Choice of the bending plane}

For applications, it is important to answer the question of which axial orientation of the crystal allows deflecting the largest number of charged particles.
To answer this question, we investigated the dependence of the efficiency of stochastic deflection mechanism as a function of the orientation of the bending plane.
Here, deflection efficiency is computed as the fraction of particles that after passage through the crystal had an angle with a bent axis lower than the critical angle of axial channeling.
For our investigation we carried out a set of simulations with 120 GeV/$c$ positrons and electrons entering a bent crystal in condition of stochastic deflection for different orientations of the bending plane. Simulations were carried out for a bent silicon crystal 2 mm thick with bending angle $\theta_B$ = 90 $\mu$rad. For each simulation, we calculated the number of particles $N$ with $(\theta_x - \theta_B)^2 + \theta_y^2 \le \psi_c^2$, where $\theta_x$ is deflection angle in the bending plane and $\theta_y$ is deflection angle in the plane that is orthogonal to the bending plane.

\begin{figure}
\includegraphics[width = \columnwidth]{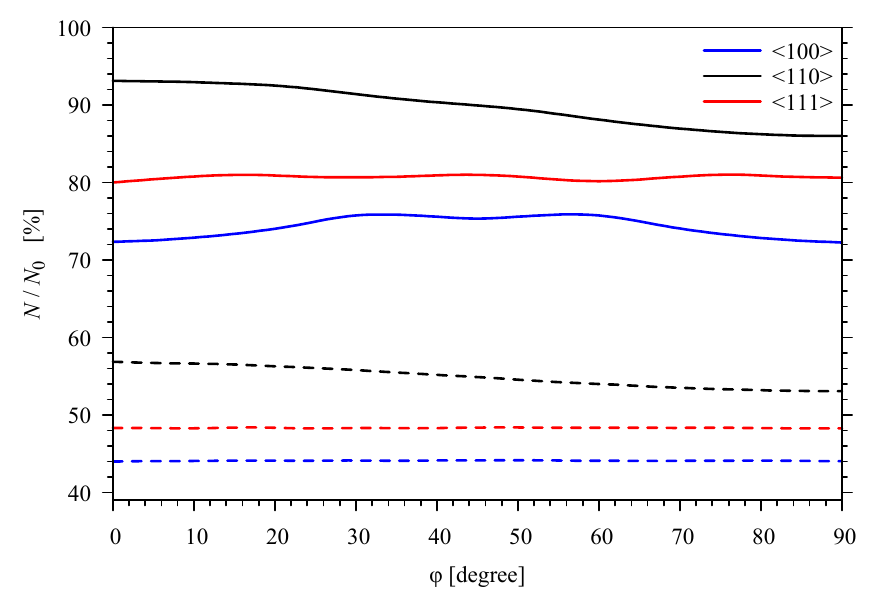}
\caption{The percentage of positrons (solid curves) and electrons (dashed curves) that remain in the regime of stochastic deflection up to the end of the crystal for different orientations of the bending plane.}
\label{fig:azimuthal}
\end{figure}

The number of particles $N_0$ that impinged on the crystal was $10^5$. In Fig. \ref{fig:azimuthal} we show the ratio $N/N_0$ for different values of the angle $\varphi$ between the bending plane and
\begin{enumerate}
  \item $(001)$ crystal plane in case of crystal orientation along $\langle 100 \rangle$ axis with respect to impinging particles (blue line);
  \item $(001)$ crystal plane in case of crystal orientation along $\langle 110 \rangle$ axis with respect to impinging particles (black line);
  \item $(\bar{1}\bar{1}2)$ crystal plane in case of crystal orientation along $\langle 111 \rangle$ axis with respect to impinging particles (red line).
\end{enumerate}
Solid curves in Fig.~\ref{fig:azimuthal} correspond to the motion of positrons in the bent crystal, while dashed curves correspond to the motion of electrons.

In Fig. \ref{fig:azimuthal}, we see that the number of positrons that remain in stochastic deflection regime up to the end of the crystal is the highest for the $\langle 110 \rangle$ orientation. One can also notice that for the $\langle 110 \rangle$ orientation the efficiency of deflection significantly depends on the angle $\varphi$. The highest efficiency in this case takes place when the bending plane coincides with the $(001)$ crystal plane (when the angle between the bending plane and the main skew planes $(1\bar{1}1)$ and $(\bar{1}11)$ is $\approx 54.7^\circ$), while the lowest efficiency corresponds to the coincidence of the bending plane with $(1\bar{1}0)$ crystal plane (when the angle of bending plane and the main skew planes is $\approx 35.3^\circ$). The difference in $N / N_0$ for $\varphi = 0$ and $\varphi = 90^\circ$ for $\langle 110 \rangle$ orientation is about 7\%.
In the experiments, the results of which were presented in Section 4, the angle $\varphi$ was equal to $\arctan{\frac{\sqrt{2}}{2}} \approx 35.3^\circ$ for the $\langle 110 \rangle$ orientation of the crystal and 0 for the $\langle 111 \rangle$ orientation.

From Eq. (\ref{eq:alp}), we see that the efficiency of stochastic deflection should be proportional to the square of the critical angle of axial channeling. This is the reason why in Fig. \ref{fig:azimuthal} the number of particles that remain in the regime of stochastic deflection up to the end of the crystal for the $\langle 111 \rangle$ orientation is smaller than for the $\langle 110 \rangle$ orientation, and in case of the $\langle 100 \rangle$ orientation it is the smallest. The squares of the critical angle of axial channeling for the orientations $\langle 110 \rangle$, $\langle 111 \rangle$, and $\langle 100 \rangle$ rate as $\sqrt{2}$:$\frac{2\sqrt{3}}{3}$:1, respectively.
Also, Eq. (\ref{eq:alp}) gives us the opportunity to say that although the numerical values of $N/N_0$ in Fig. \ref{fig:azimuthal} will change with a change in energy, the figure will not change qualitatively. Axial orientation $\langle 110 \rangle$ will remain the most successful for deflecting the largest number of charged particles.

Comparing solid and dashed curves in Fig. \ref{fig:azimuthal}, we see that in case of negatively charged particles the efficiency of deflection is smaller, because of the higher intensity of incoherent scattering, which leads to the escape from the stochastic deflection regime. In contrast to the case of positively charged particles, for electrons we see that the dependence of the deflection efficiency on the angle $\varphi$ is present only in the case of $\langle 110 \rangle$ orientation of the crystal. The difference in $N / N_0$ for $\varphi = 0$ and $\varphi = 90^\circ$ in this case is about 4\%.

\section{Conclusions}

An investigation on the mechanism of stochastic deflection of axially confined ultra-high energy charged particles in bent crystals was carried out.
We tested the stochastic deflection mechanism using electrons and positrons with an experiment carried out in the H4 beam line of the CERN Super Proton Synchrotron (SPS) - North Area, where electron and positron beams of 120 GeV/$c$ were available.
Indeed, this was the first experimental evidence of axial deflection of ultrarelativistic electrons and positrons (or in general leptons) in bent crystals.
The investigation showed that stochastic deflection is efficient for deflection of both positively and negatively charged particles. It was also shown that stochastic deflection is more efficient for the crystal orientation near axis with higher atomic string potential (in case of a silicon crystal it is the $\langle 110 \rangle$ axis). Moreover, we have shown that for the $\langle 110 \rangle$ orientation of the Si crystal, the efficiency of deflection is highly dependent on the choice of the bending plane for both positively and negatively charged particles.
Although the experimental results and simulation results presented in the article were obtained for particles with a momentum of 120 GeV/$c$, Eq. (\ref{eq:alp}) allows us to conclude that the axial orientation $\langle 110 \rangle$ is the most successful for deflecting the largest number of charged particles also at other momentum values.
At the same time, due to the more intense incoherent scattering, the denominator in Eq. (\ref{eq:alp}) for negatively charged particles is larger than for positively charged ones. The consequence of this is the greater efficiency of deflecting positively charged particles as compared to negatively charged ones.
Nevertheless, negative particle deflection under axial alignment is more efficient than for planar channeling (see for instance \cite{Bagli2017}).
The presented results are relevant for a possible application of axial deflection, with the advantage of larger efficiency than for the planar case, for extraction/\hspace{0cm}collimation of particles in the next-generation of particle accelerators/\hspace{0cm}colliders, where positive and negative lepton beams will be involved, such as FCC-ee, muon colliders and linear colliders.
For collimation in linear colliders and muon colliders, it is required to steer both positive and negative beams. Since planar deflection could be highly inefficient to steer TeV negative beams, axial deflection could be really a good option.
Indeed, as shown in \cite{shul2013stochastic}, at a particle energy of 1 TeV, the stochastic deflection makes it possible to deflect almost 100\% of the beam particles in the direction of the crystal bending for both positive and negative particles, since incoherent scattering is strongly suppressed for such a high energy.
So the stochastic deflection can be applied in future ultrahigh-energy colliders.

\section{acknowledgements}
This work was partially supported by the INFN through the ELIOT experiment and the OSCaR project and by the National Academy of Sciences of Ukraine (budget program ``Support for the Development of Priority Areas of Scientific Research'' (6541230), projects C-2/50-2020 and F30-2020). We acknowledge the support of the PS/SPS physics coordinator and of the CERN SPS, the EN-EA group technical staff. M. Romagnoni acknowledges support from the ERC Consolidator Grant SELDOM G.A. 771642.

\bibliographystyle{spphys}
\bibliography{biblio}

\begin{thebibliography}{10}
\providecommand{\url}[1]{{#1}}
\providecommand{\urlprefix}{URL }
\expandafter\ifx\csname urlstyle\endcsname\relax
  \providecommand{\doi}[1]{DOI \discretionary{}{}{}#1}\else
  \providecommand{\doi}{DOI \discretionary{}{}{}\begingroup
  \urlstyle{rm}\Url}\fi

\bibitem{biryukov2013crystal}
V.M. Biryukov, Y.A. Chesnokov, V.I. Kotov, \emph{Crystal channeling and its
  application at high-energy accelerators} (Springer Science \& Business Media,
  2013)

\bibitem{IvanovVolume}
Y.M. Ivanov, A.A. Petrunin, V.V. Skorobogatov, et~al., Phys. Rev. Lett.
  \textbf{97}, 144801 (2006)

\bibitem{afonin2012observation}
A.G. Afonin, V.T. Baranov, M.K. Bulgakov, et~al., Physical Review Special
  Topics-Accelerators and Beams \textbf{15}(8), 081001 (2012)

\bibitem{ScandaleAxial}
W.~Scandale, A.~Vomiero, S.~Baricordi, et~al., Phys. Rev. Lett. \textbf{101},
  164801 (2008)

\bibitem{Scandale2016129}
W.~Scandale, G.~Arduini, M.~Butcher, et~al., Phys. Lett. B \textbf{758}, 129
  (2016)

\bibitem{splitting1}
L.~Bandiera, I.~Kirillin, E.~Bagli, et~al., Nucl. Instrum. Methods Phys. Res. B
  \textbf{402}, 296 (2017)

\bibitem{splitting2}
L.~Bandiera, A.~Mazzolari, E.~Bagli, et~al., Eur. Phys. J. C \textbf{76}(2), 1
  (2016)

\bibitem{Lin}
J.~Lindhard, Mat. Fys. Medd. Dan. Vid. Selsk \textbf{34}, 1 (1965)

\bibitem{Akhiezer:Shulga}
A.I. Akhiezer, N.F. Shulga, \emph{{High-energy Electrodynamics in Matter}}
  (Gordon $\&$ Breach, New York, 1996)

\bibitem{ScandaleQMsps}
W.~Scandale, G.~Arduini, R.~Assmann, et~al., Phys. Lett. B \textbf{692}(2), 78
  (2010)

\bibitem{Afonin1998}
A.G. Afonin, V.M. Biryukov, V.A. Gavrilushkin, et~al., JETP Lett. \textbf{67},
  781 (1998)

\bibitem{Lanzoni2010}
V.~Guidi, L.~Lanzoni, A.~Mazzolari, J. Appl. Phys. \textbf{107}(11), 113534
  (2010)

\bibitem{Shulga1995}
N.~Shul'ga, A.~Greenenko, Phys. Lett. B \textbf{353}, 373  (1995)

\bibitem{Grinenko1991}
A.A. Grinenko, N.F. Shul'ga, Sov. JEPT Lett. \textbf{54}, 524 (1991)

\bibitem{SHULGA2011100}
N.~Shul'ga, I.~Kirillin, V.~Truten', Phys. Lett. B \textbf{702}(1), 100  (2011)

\bibitem{Bak1984}
J.~Bak, P.~Jensen, H.~Madsb\o{}ll, et~al., Nucl. Phys. B \textbf{242}(1), 1
  (1984)

\bibitem{Baurichter1986}
A.~Baurichter, K.~Kirsebom, R.~Medenwaldt, et~al., Nucl. Instrum. Methods Phys.
  Res. B \textbf{119}(1), 172  (1996)

\bibitem{SCANDALE2016826}
W.~Scandale, G.~Arduini, M.~Butcher, et~al., Phys. Lett. B \textbf{760}, 826
  (2016)

\bibitem{garattini-epjc}
W.~Scandale, F.~Andrisani, G.~Arduini, et~al., Eur. Phys. J. C \textbf{78}, 505
  (2018)

\bibitem{Scandale2009301}
W.~Scandale, A.~Vomiero, E.~Bagli, et~al., Phys. Lett. B \textbf{680}, 301
  (2009)

\bibitem{GERMOGLI201581}
G.~Germogli, A.~Mazzolari, L.~Bandiera, E.~Bagli, V.~Guidi, Nucl. Instrum.
  Methods Phys. Res. B \textbf{355}, 81  (2015)

\bibitem{Ban3}
L.~Bandiera, E.~Bagli, G.~Germogli, et~al., Phys. Rev. Lett. \textbf{115},
  025504 (2015)

\bibitem{PhysRevLett.112.135503}
A.~Mazzolari, E.~Bagli, L.~Bandiera, et~al., Phys. Rev. Lett. \textbf{112},
  135503 (2014)

\bibitem{SYTOV2017}
A.~Sytov, L.~Bandiera, D.~De~Salvador, et~al., Eur. Phys. J. C \textbf{77}, 901
  (2017)

\bibitem{Bagli2017}
E.~Bagli, V.~Guidi, A.~Mazzolari, et~al., Eur. Phys. J. C \textbf{77}(2) (2017)

\bibitem{Lie}
D.~Lietti, E.~Bagli, S.~Baricordi, et~al., Nucl. Instrum. Methods Phys. Res. B
  \textbf{283}, 84 (2012)

\bibitem{Wistisen2017}
T.~Wistisen, R.~Mikkelsen, U.~Uggerh\o{}j, et~al., Phys. Rev. Lett.
  \textbf{119}(2) (2017)

\bibitem{PhysRevLett.114.074801}
U.~Wienands, T.W. Markiewicz, J.~Nelson, et~al., Phys. Rev. Lett. \textbf{114},
  074801 (2015)

\bibitem{wistisen2016channeling}
T.~Wistisen, U.~Uggerh{\o}j, U.~Wienands, et~al., Physical Review Accelerators
  and Beams \textbf{19}(7), 071001 (2016)

\bibitem{ban}
L.~Bandiera, E.~Bagli, V.~Guidi, et~al., Phys. Rev. Lett. \textbf{111}, 255502
  (2013)

\bibitem{Kirillin}
I.V. Kirillin, N.F. Shul'ga, L.~Bandiera, V.~Guidi, A.~Mazzolari, Eur. Phys. J.
  C \textbf{77}, 117 (2017)

\bibitem{kirillin_dependence_2017-1}
I.V. Kirillin, Probl. Atomic Sci. Technol. \textbf{109}(3), 67 (2017)

\bibitem{Relaxation}
L.~Bandiera, A.~Mazzolari, E.~Bagli, et~al., Eur. Phys. J C \textbf{76}, 1
  (2016)

\bibitem{optimal}
I.V. Kirillin, Phys. Rev. Accel. Beams \textbf{20}(10), 104401 (2017)

\bibitem{chesnokov_about_2014}
Y.A. Chesnokov, I.V. Kirillin, W.~Scandale, et~al., Phys. Lett. B \textbf{731},
  118 (2014)

\bibitem{kirillin_orientation_2015}
I.V. Kirillin, N.F. Shul’ga, Nucl. Instrum. Meth. B \textbf{355}, 49 (2015)

\bibitem{kirillin_dependence_2017}
I.V. Kirillin, N.F. Shul’ga, Nucl. Instrum. Meth. B \textbf{402}, 40 (2017)

\bibitem{NA-beam}
\urlprefix\url{https://sba.web.cern.ch/sba/BeamsAndAreas/H2/H2_presentation.html}

\bibitem{Baricordi2007}
S.~Baricordi, V.~Guidi, A.~Mazzolari, et~al., Appl. Phys. Lett. \textbf{91}(6),
  061908 (2007)

\bibitem{baricordi2008}
S.~Baricordi, V.~Guidi, A.~Mazzolari, D.~Vincenzi, M.~Ferroni, J. Phys. D:
  Appl. Phys. \textbf{41}(24), 245501 (2008)

\bibitem{ROSSI2015369}
R.~Rossi, G.~Cavoto, D.~Mirarchi, S.~Redaelli, W.~Scandale, Nucl. Instrum.
  Meth. B \textbf{355}, 369 (2015)

\bibitem{Lietti2013}
D.~Lietti, A.~Berra, M.~Prest, E.~Vallazza, Nucl. Instrum. Methods Phys. Res. A
  \textbf{729}, 527  (2013)

\bibitem{telescope}
\urlprefix\url{https://cds.cern.ch/record/1353904/files/Thesis-2011-Hasan.pdf}

\bibitem{Bagli2017independence}
E.~Bagli, et~al., Eur. Phys. J. C \textbf{77}(2), 71 (2017)

\bibitem{shul2013stochastic}
N.F. Shul'ga, I.V. Kirillin, V.I. Truten', J. Surf. Invest.: X-Ray, Synchrotron
  Neutron Tech. \textbf{7}(2), 398 (2013)

\end{thebibliography}

\end{document}